\documentclass[aip,apl,reprint]{revtex4-1}
\usepackage[active]{srcltx}
\usepackage{graphicx}
\usepackage{amsmath}
\usepackage{mathcomp}	
\begin{document}
\title{Joule-assisted silicidation for short-channel silicon nanowire devices}

\author{Massimo Mongillo}
\altaffiliation{Present adress: IMEP-LAHC, Grenoble-INP, MINATEC, 3 Parvis Louis Neel, B.P. 257, 38016 Grenoble, France, E-mail: massimo.mongillo@minatec.inpg.fr}
\affiliation{SPSMS/LaTEQS, CEA-INAC/UJF-Grenoble 1, 17 Rue des Martyrs, 38054 Grenoble Cedex 9, France}
\author{Panayotis Spathis}
\affiliation{SPSMS/LaTEQS, CEA-INAC/UJF-Grenoble 1, 17 Rue des Martyrs, 38054 Grenoble Cedex 9, France}
\author{Georgios Katsaros}
\affiliation{SPSMS/LaTEQS, CEA-INAC/UJF-Grenoble 1, 17 Rue des Martyrs, 38054 Grenoble Cedex 9, France}
\author{Pascal Gentile}
\affiliation{SP2M/SINAPS, CEA-INAC/UJF-Grenoble 1, 17 Rue des Martyrs, 38054 Grenoble Cedex 9, France}
\author{Marc Sanquer}
\author{Silvano De Franceschi}
\affiliation{SPSMS/LaTEQS, CEA-INAC/UJF-Grenoble 1, 17 Rue des Martyrs, 38054 Grenoble Cedex 9, France}
\email{silvano.defranceschi@cea.fr}

\begin{abstract}

We report on a technique enabling electrical control of the contact silicidation process in silicon nanowire devices. Undoped silicon nanowires were contacted by pairs of nickel electrodes and each contact was selectively silicided by means of the Joule effect. By a realtime monitoring of the nanowire electrical resistance during the contact silicidation process we were able to fabricate nickel-silicide/silicon/nickel- silicide devices with controlled silicon channel length down to 8 nm.
\end{abstract}


\maketitle

The performance and production costs of computer microprocessors have been improving over time according to an exponential trend commonly known as the Moore's law. This evolution has been possible through a progressive downward scaling in the size of silicon complementary metal-oxide-semiconductor (CMOS) transistors, the elementary building blocks of microprocessors. According to the International Technology Roadmap for Semiconductors the transistor gate length, which defines the device active channel, is expected to approach 20 nm in a matter of a few years \cite{ITRS}. Below such a length scale, however, a diffused concern exists that the conventional transistor paradigm may no longer provide a viable option.  Multiple issues could contribute to this breakdown, such as an increasingly large gate leakage current or limited doping control (in short channel transistors, variations in the spatial distribution of doping impurities can give rise to large statistical variability in the performance of transistors or even degrade the on-off switching characteristic \cite{Roy2006}). On the other hand, the extreme miniaturization of silicon transistors may open the way to new operating principles relying on quantum mechanical effects associated with tunneling and size confinement. Along this line, a variety of novel device concepts have been proposed such as Schottky-barrier transistors \cite{Larson2006},  band-to-band tunneling transistors \cite{Bjork2008}, resonant tunneling transistors \cite{Reed1989}, and also single-electron transistors \cite{Fulton1987} relying on the Coulomb blockade effect, a phenomenon that has been found to survive up to room temperature in ultra-small silicon devices \cite{Shin2010}. 

The implementation and testing of these innovative device concepts require the characteristic device sizes, in particular the channel length, to be defined with nm-scale precision. In typical silicon transistors, fabricated by conventional top-down techniques, short-channel lengths are achieved by selective-area dopant implantation using the transistor gate electrode as a self-aligned mask. The fabrication of such devices, however, is rather cumbersome due to the large number of processing steps. An alternative route has recently become accessible thanks to the development of chemically synthesized Si nanowires (NWs). Such silicon NWs attract widespread interest in view of their use as building blocks for the bottom-up assembly of a variety of devices with potential applications in low-cost electronics, biochemical sensors, and solar cells ~\cite{McAlpine2003,Cui2001a,Tian2007}. Owing to the relative simplicity and versatility of the bottom-up fabrication approach, Si NWs provide at the same time a practical test bench for new device concepts. 

Following a groundbreaking work by Wu \cite{Wu2004}, short-channel transistors were recently fabricated ~\cite{Weber2006,Appenzeller2006,Hu2008,Lin2008,Zwanenburg2009} from individual Si NWs by taking advantage of the so-called silicidation process, a thermally-activated solid-state reaction where a metal M (where M = Ni, Pt, Co, \textit{etc}.) diffuses into the silicon creating different metallic phases such as MSi, M$_2$Si, MSi$_2$, \textit{etc}. Metal silicides are routinely used in conventional transistors in view of their ability to lower the contact resistance between the silicon channel and the metal source/drain leads, thereby reducing dissipation and on-off switching times \cite{ITRS}. 
It has been shown that upon thermal annealing a metal silicide can extend along the NW axis over micron-scale distances from a metal contact electrode. This has enabled the fabrication of NW transistors with channel lengths down to a few tens of nm ~\cite{Hu2008,Zwanenburg2009} (see also Refs  ~\cite{Schmitt2010,Chou2010} for a recent review). The Si NWs were individually contacted by pairs of metal electrodes with micron-scale separation and the channel length was set by the final distance between the silicide sections originating from the two contact electrodes.

In all of these works, the metal silicides were formed by conventional thermal annealing techniques, \textit{i.e.} by heating the entire device substrate to temperatures ranging between 280 and 550~$^\circ$C. This type of approach allows only a relatively poor control of the device channel length which can be affected by several factors such as the NW diameter~\cite{Appenzeller2006}, the distance between the lithographically-defined metal electrodes and the sensitivity of the silicide formation to the surface chemistry \cite{Lee2000}. The variability in these parameters makes the realization of short-channel devices a low-yield statistical process requiring a post-processing selection of the short-channel devices, usually done by electron-beam imaging. Alternatively, Lu ~\cite{Lu2007} showed that thermal annealing and high-resolution imaging could be simultaneously performed in a transmission electron microscope. This approach was shown to provide remarkably short and well-controlled Si sections between the two silicide phases. Since these heterostructures were realized in NWs with unconnected metal leads, however, no actual devices were in fact demonstrated in that work.

\begin{figure}[t]
\includegraphics[width=8.5cm]{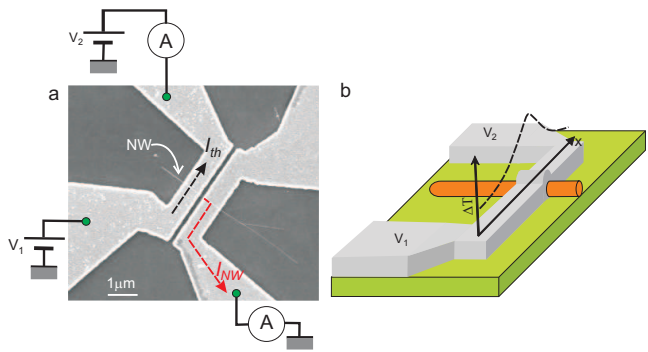}
 \caption{Illustration of the Joule-assisted silicidation technique. (a) SEM image of a representative device fabricated from a single Si NW. The two Ni contacts are annealed one at a time by flowing a heating current $I_{th}$, driven by the voltage difference $V_{1}-V_{2}$, and simultaneously measuring the current $I_{NW}$ through the Si NW, driven by the voltage average $(V_{1} + V_{2})/2$. The shown circuit diagram refers to the configuration used for the annealing of the left contact. (b) Qualitative temperature profile along a nickel strip as a result of Joule heating. Because of the heat losses toward the substrate and toward the wider Ni lines at the edges of the strip, a temperature gradient builds up along the strip with a temperature peak at the strip mid point.  }

 \label{Fig1}%
 \end{figure}

In this Article we present a  technique enabling the fabrication of well-controlled short-channel silicide/silicon/silicide tunnel devices from undoped Si NWs. The thermal energy required for the silicidation process is generated by the flow of an electrical current through the metal contact and the resulting Joule effect. The key feature of our technique lies in the fact that the protrusion of the metal silicide into the NW is controlled in real time by monitoring the device resistance during the silicidation process. 
\section{Results and discussion}
We used undoped Si NWs with diameters in the $20 - 30$ nm range and lenghts of several microns. The Si Nws were individually contacted by pairs of parallel Ni strips as shown in Fig.\ref{Fig1}(a). The two Ni strips were positioned in such a way to have the contacted Si Nw crossing both of them around their mid-point as shown qualitatively in Fig.\ref{Fig1}(b).
This configuration was chosen in order to optimize the contact annealing process, since Joule heating induces a nonuniform temperature profile along the strip with a maximum at its mid-point as shown in Fig.\ref{Fig1}(b).

The two contacts were annealed one at a time. For the annealing of the left contact, two independently controlled voltages, $V_1$ and $V_2$, were applied to the left strip as shown in  Fig.\ref{Fig1}(a). The difference $V_{th}$ between these voltages was used to induce the heating electrical current $I_{th}= V_{th}/R_{\mathrm{strip}}$ along the strip, with $R_{\mathrm{strip}}$ being the strip resistance (for simplicity we have neglected the series resistance of the wider connecting lines).

The average voltage between  $V_1$ and $V_2$ was adjusted to define the electrostatic potential $V_{NW} \approx (V_1 + V_2)/2 $ at the contact between the left metal strip and the NW. Connecting one edge of the right metal strip to ground \textit{via} an amperometer enabled the current $I_{NW} $ through the NW to be monitored in real time during the silicidation process. This current, induced by the applied voltage $V_{NW} $, provided a measurable feedback parameter to control the silicidation. The same procedure was repeated for the annealing of the right contact after swapping the electrical connections to the nickel strips. The effect of the back gate voltage was found to be unimportant due to the poor capacitive coupling between the gate and the NW channel. Hence the back gate was kept at ground throughout all of the experiments discussed further below.

\begin{figure}
\includegraphics[width=8.5cm]{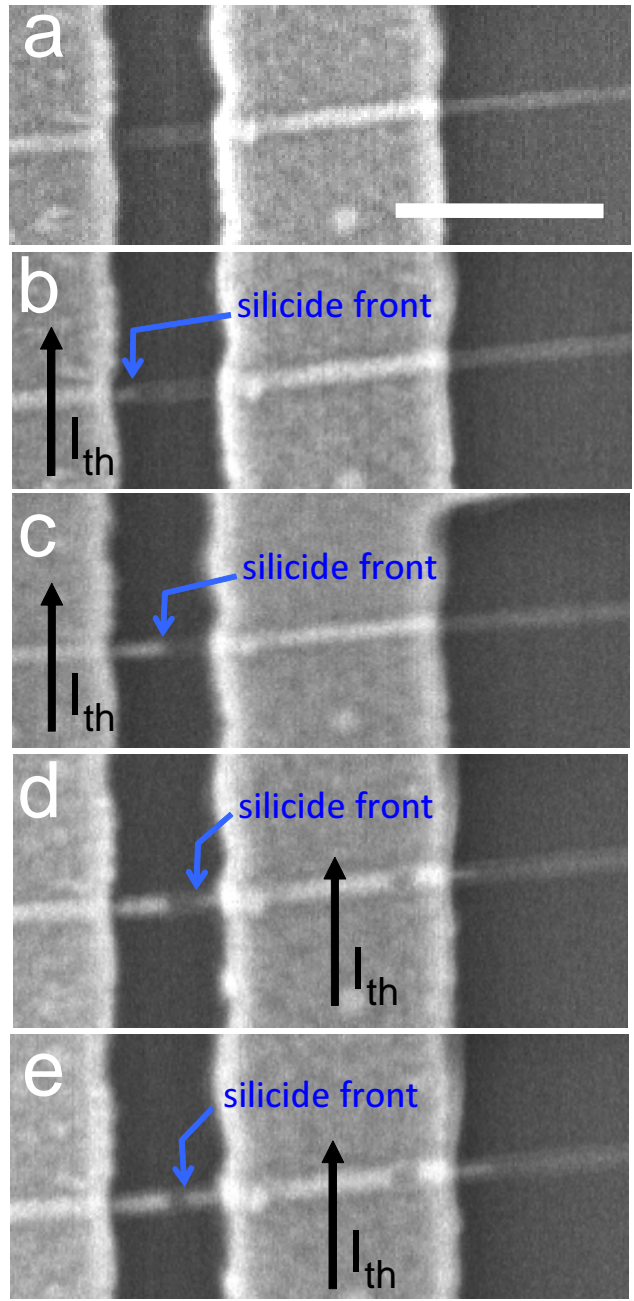}%
 \caption{SEM close-ups of a device during the Joule-assisted silicidation of the nickel contacts. (a) SEM image of the fabricated device before annealing. Scale bar: 500 nm. (b)-(c) SEM Images showing the protrusion of the silicide phase from the left nickel strip. During these images the heating current,  through the left strip was 18 and 20 mA, respectively. (d)-(e) Both SEM images were taken for the same $I_{th} = 20$ mA with a delay of 6 minutes between the first (d) and the second snapshot (e). Eventually a silicide/silicon/silicide device with a channel length of about 50 nm was obtained (e). Intermediate snapshots are given as Supplementary Information. } %
 \label{Fig2}
 \end{figure}

For a first feasibility test of the Joule-assisted silicidation technique, an initial set of experiments were carried out inside a scanning electron microscope (SEM), a Hitachi 4100S equipped with electrical feedthroughs. Due to the interaction with the electron beam of the microscope, however, a proper measurement of $I_{NW} $ could not be performed while taking SEM images. 
Fig.\ref{Fig2} shows a sequence of such images taken during the Joule-assisted annealing procedure for a representative device. 
$V_{th} $ and, correspondingly, $I_{th} $ were first applied to the left metal strip and gradually increased in small steps while continuously taking SEM snapshots of the NW device. No effect was observed up to $I_{th} \approx 18$ mA. Above this threshold, a nickel silicide phase was seen to emerge from the heated contact as shown in  Fig.\ref{Fig2}(b) and Fig.\ref{Fig2}(c), where the silicide phase can be easily distinguished from pure Si by virtue of its brighter contrast. This demonstrates the effectiveness of Joule annealing to promote the silicidation process. After having diffused into the NW section below the contact, the Ni atoms begin to diffuse along the NW longitudinal axis, forming a nickel silicide phase protruding from the metal strip. The SEM images show as well that the silicon/silicide interface remained sharply defined all along the silicidation process, consistent with earlier reports where standard annealing techniques had been employed \cite{Wu2004,Weber2006}. 

The heating current was increased up to $I_{th}=20$~mA corresponding to a current density $J_{th} \approx 3\times 10^{11}$ A/m$^{2}$, close to the maximum current density typically achieved in these nickel strips before the observation of an electrical breakdown due to electromigration  ($J^{max}_{th}\approx 4\times 10^{11}$A/m$^{2}$).    
The silicide phase was found to protrude only up to a distance of $150-200$ nm from the heated contact. This self-limited protrusion can be ascribed to the locality of the generated heat and to the substrate acting as a thermal sink. In fact, as the silicide front moves away from the heated contact, it cools down due to heat losses towards the substrate. Its progression ceases when the temperature at the silicide/silicon interface becomes too low to further promote the diffusion of the Ni atoms into the Si NW. We expect that, during the Joule-assisted silicidation, the temperature at the silicide/silicon interface is close to the minimum temperature required for silicide formation ($\sim 250^{\circ}$C)\cite{Foggiato2004}. As a result, the most favourable silicide phase near the silicide/silicon is presumably Ni$_{2}$Si.

The heating voltage $V_{th} $ was successively applied to the right metal strip. The corresponding Joule annealing process was deliberately interrupted before reaching the complete silicidation of the NW segment between the contacts. In this way, a short Si channel could be formed between the silicide sections as shown in   Fig.\ref{Fig2}(e). 
Besides enabling some degree of control over the device channel length, real-time SEM imaging allowed us to identify the characteristic time scales underlying the Joule-assisted silicidation process. In fact,  Fig.\ref{Fig2}(d) and Fig.\ref{Fig2}(e) were recorded with a delay of six minutes at a constant value of the heating current ($I_{th}=20$~mA). More SEM images were also taken between these two snapshots in order to follow the intermediate evolution of the silicidation process (see Supporting Information). The acquisition of each SEM image required about 20 s. From the sequence of SEM images we inferred that, following the last step in $I_{th}$, the silicide front moved at a velocity of the order of 10 nm/min. This relatively slow dynamics imposes no stringent requirements on the rapidity of a feedback control on the silicidation process.

\begin{figure}
\includegraphics[width=8.5cm]{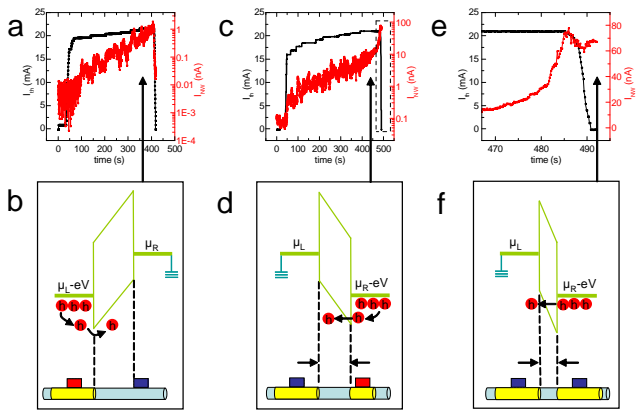}%
 \caption{\label{Fig3} Example of the experimental procedure to obtain a silicide/silicon/silicide tunnel device by means of the Joule annealing technique. 
(a) Silicidation of the left Ni contact. The heating current, $I_{th}$, through the left Ni strip and the simultaneously measured nanowire current, $I_{NW}$, are shown as a function time. The step-by-step increase of $I_{th}$ (black trace) is followed by an increase in $I_{NW}$ denoting the formation of a nickel silicide phase. (b) Qualitative band diagram of the Si nanowire device at an intermediate stage (denoted by the black arrow) where the Ni silicide (shown in yellow) has protruded into the silicon nanowire (in light blue). A linear band profile is assumed in accordance with the low background doping level, $p\sim 10^{15} - 10^{16}$ cm$^{-3} $. The dominant transport mechanism responsible for $I_{NW}$, \textit{i.e.} the thermionic emission of holes over the positively-biased silicide/silicon Schottky contact, is schematically shown. (c)-(d) Same as (a)-(b), but now for the silicidation of the right contact. The electric-field enhancement following from the shrinking channel length leads to an increasing contribution from thermally-assisted tunneling across the edge of the right Schottky barrier. (e)-(f) Close-up of the region delimited by a black dashed rectangle in (c) and corresponding band diagram. $I_{NW}$ exhibits a steep increase up to 75 nA, while $I_{th}$ is kept constant and equal to 21 mA. When $I_{th}$ is brought to zero, $I_{NW}$ does not drop, contrary to what observed in (a). This behavior denotes the onset of temperature-independent direct tunneling through the 8-nm Si section between silicide contacts as schematically represented in (f).}
\end{figure}

To allow the implementation of  such a feedback control, based on the measurement of $I_{NW}$, the Joule annealing experiments were eventually performed in a vacuum chamber ($\sim 10^{-6}$ mbar) at ambient temperature and in dark conditions. Performing the silicidation in vacuum prevented the oxidation of the Ni strips and of the Si NW.
We illustrate the Joule annealing technique with an example on a representative device.  To begin with, $V_1$ and $V_2$ were applied to the left nickel strip, and both of them were initially set to 1 V in order to impose a bias voltage $V_{NW}=1$ V across the Si NW. Because as-deposited contacts have very high resistances (exceeding $10^{11}  \Omega$), the resulting current was found to be below the measurement noise level as shown in  Fig.\ref{Fig3}(a).
As for the previously discussed experiments,  $V_{th}$ (and, with it, $I_{th}$) was increased from 0 to typical voltages of $\sim$1.5 V, in small ($\sim$50 mV) steps 
(the precise magnitude of these steps was found to be relatively unimportant, especially during the initial phase of the Joule annealing process).
Meanwhile, $(V_1 + V_2)/2$ was kept equal to 1 V in order to maintain an approximately constant bias across the Si NW.
A measurable increase in the NW current was observed for $I_{th} > 19$ mA denoting the onset of silicidation. Above this threshold, $I_{NW}$ was found to increase with $I_{th}$ due to the protrusion of the silicide phase along the NW axis. At $I_{NW}\sim$1 nA, the heating current was turned off by abruptly bringing $V_{th}$ to zero and $I_{NW}$ quickly dropped to about 30 pA, \textit{i.e.} just above the instrumental noise level. This dramatic drop is a clear signature of thermally-activated transport, with a predominant contribution from the thermionic emission of holes over the Schottky barrier at the silicide/silicon interface ( Fig.\ref{Fig3}(b)). In fact, following the removal of the heating current, temperature returned back to ambient conditions leading to the observed drop in $I_{NW}$.

After the first annealing step the roles of the two metal strips were exchanged to promote the silicidation of the right contact ( Fig.\ref{Fig3}(c)). As for the
first annealing process, we set $V_{NW} \approx (V_1 + V_2)/2 = 1$ V and increased $V_{th}$  in small steps. After the last step, \textit{i.e.} for $I_{th} = 21$ mA, $I_{NW}$ began to rise steeply with time up to about 75 nA. At that stage, $V_{th}$ was abruptly brought back to zero to prevent $I_{NW}$ from growing even further and the Si NW channel from being entirely converted into silicide. Contrary to what observed in the annealing of the left contact ( Fig.\ref{Fig3}(a)), following the removal of the heating current, $I_{NW}$ remained roughly constant denoting a clear insensitivity to temperature ( Fig.\ref{Fig3}(e)). We explain this experimental finding as follows.
At the beginning of the second annealing process, $I_{NW}$ is essentially dominated by the thermionic emission of holes over the
reversely-biased Schottky barrier at the right silicon/silicide interface ( Fig.\ref{Fig3}(d)), this transport mechanism being favored by the locally enhanced
temperature. As the silicon/silicide interface approaches the other silicide/silicon interface, previously formed from the silicidation of the left contact, the electric field in the silicon channel,
$F= V_{NW}/L$,  increases as a result of the shrinking channel length, $L$. This electric field leads to an
increasing contribution from thermally-assisted tunneling processes across the triangular-shaped Schottky barrier. Consequently, the effective barrier height lowers as $L$ gets shorter. In the limit of very short channel lengths, thermal activation becomes no longer necessary to promote tunneling and transport. This limit is reached at the end of the annealing process shown in  Fig.\ref{Fig3}(f). According to the Fowler-Nordheim model, $I_{NW}$ is expected to grow proportionally to $L^{-2}$exp$(-\alpha \times L )$, with $\alpha$ being a fixed parameter that depends on material and geometrical properties. This explains the observed current shoot-up and the fact that $I_{NW}$ does not drop when Joule heating is switched off. 

\begin{figure}
\includegraphics[width=8.5cm]{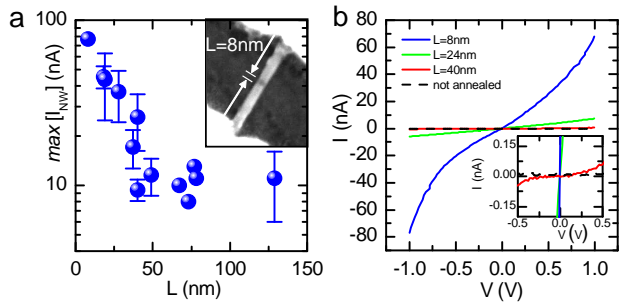}%
 \caption{\label{Fig4} Relation between transport properties and device channel length. (a) The maximum value of the nanowire current, $max[I_{NW}]$, measured during the annealing process of the second contact, is plotted against the corresponding channel length, $L$, measured by SEM after the completion of Joule annealing procedure. A clear correlation is found for $L < 50$ nm, with $max[I_{NW}]$ increasing for smaller $L$. This demonstrates that, in the limit of relatively short channels, $max[I_{NW}]$ can be used as a feedback parameter to achieve a desired $L$. 
 Vertical error bars correspond to the standard deviation for $max[I_{NW}]$ as a function of time. Horizontal error bars ($\pm 5$ nm) are equal to the dot size.
For a few devices, the silicidation process was stopped for $max[I_{NW}]$ exceeding 100 nA. In these cases, however, $L$ was too small to be measured by SEM.
Inset: SEM image of a short-channel device. The corresponding Joule annealing process was reported in  Fig.\ref{Fig3}. (b) Room-temperature, current-voltage ($I-V$) characteristics of a representative ensemble of nanowire devices before (black trace) and after (red, blue, green traces) the Joule-assisted silicidation of the contacts. At higher bias voltages nonlinearities can be observed as a result of Fowler-Nordheim tunneling (the tunnel barrier becomes triangular). Inset: close-up of the IV characteristics emphasizing the difference between the 40-nm-channel device and the not-annealed device. }%
\label{fig4}
 \end{figure}

The Joule annealing technique illustrated in  Fig.\ref{Fig3} was applied to 15 Si NW devices. Each time, the maximum NW
current, $max[I_{NW}]$, reached during the second annealing process and measured just before switching off the heating current, was varied in order to obtain devices with different channel lengths, measured a posteriori by SEM. The correlation found between $L$ and $max[I_{NW}]$ is shown in   Fig.\ref{Fig4}(a). 
In the limit of relatively large $L$ (\textit{i.e.} for $L \geq 50$nm) the measured $max[I_{NW}]$ is dominated by the thermionic emission of
holes over the reverse-biased Schottky barrier at the 'heated' contact. The effective barrier lowering due to tunneling has little relevance in this case which explains why $max[I_{NW}]$ is essentially independent of $L$. For $L <  50$ nm, however, the contribution from
Fowler-Nordheim tunneling becomes important and $max[I_{NW}]$ increases steeply when $L$ gets shorter. Therefore, the data points for $L <  50$ nm provide a calibration curve that can be used to fabricate short-channel devices with well-controlled channel length. This fabrication approach does not require real-time SEM imaging. Moreover it is of much simpler implementation and it can yield extremely short channel lengths hardly measurable by SEM. In fact we were able to reproducibly obtain NW devices with channel lengths below 10 nm. One of such devices is shown in the inset of   Fig.\ref{Fig4}(a). We estimate a silicon channel length of $8 \pm 5$ nm, where the uncertainty comes mainly from SEM resolution. 

The current-voltage ($I-V$) characteristics of all devices were measured both before and after the Joule-assisted annealing of the contacts. Some representative results are reported in  Fig.\ref{fig4}(b). Before the annealing, all of the devices basically showed no conduction, with resistances exceeding $10^{11} \Omega$.    
Annealed devices with relatively long silicon channel exhibited measurable conduction only for bias voltages larger than a few hundred mV. This is the case of the red trace in  Fig.\ref{fig4}(b) which was obtained for a device with a 40-nm-long silicon channel. The onset of conduction at relative large bias voltages can be ascribed to the electric-field enhancement of tunneling processes, which are dominated by hole tunneling through the silicon/silicide Schottky barrier.  
In the case of short-channel devices, on the other hand, tunneling was found to occur also for small bias voltages leading to almost linear characteristics with zero-bias resistances decreasing with $L$ (see the inset of Fig.\ref{fig4}(b)). Representative I-V characteristics are shown    for $L = 8$ and 24 nm, with corresponding resistances of 22 and 160 M$\Omega$, respectively. 
\section{Conclusions}
We have demonstrated a  technique based on Joule-assisted silicidation allowing the realization of well-controlled tunnel devices from single Si NWs. Although this technique has been implemented using undoped Si NWs and nickel contacts, we believe it could be readily applied also to doped Si NWs.  In the presence of doping, the protrusion of the metal silicide into the nanowire is expected to push away the doping impurities forcing them to pile up next to the silicide front edge \cite{Qiu2008}. This phenomenon can produce a significant narrowing of the Schottky barrier such that relatively low device resistances can be achieved even for relatively long channel lengths. The capability of our technique to controllably attain device channel lengths below 10 nm could as well open an interesting route towards short-channel, single-dopant devices whose operation stem from resonant tunneling through individual doping impurities. Finally, our Joule annealing technique could be used for any type of contacting metal capable of forming a silicide phase. For instance, we have tested our technique also for Pt contacts to Si NWs and short-channel PtSi/Si/PtSi heterostructures could be obtained. The versatility of the Joule annealing technique may yield new control over a wide class of devices such as 
Ge/Si core-shell structures \cite{Hu2008}  with NiSi$_{x}$Ge$_{y}$ contacts, high-mobility Ge channels \cite{Burchhart2009} with copper germanide contacts; Si NWs with manganese silicide contacts \cite{Lin2010} (possibly relevant for spintronics), or cobalt-silicide contacts (relevant also for hybrid superconductor-semiconductor devices).  We further believe that our technique could as well be applied to other semiconductor nanostructures (\textit{e.g.} III-V semiconductor nanowires) for which contact fabrication requires a thermal alloying process (\textit{e.g.} AuGe contacts to GaAs).

\section{Experimental section}

\textbf{Device fabrication}. The Si NWs were grown by chemical vapor deposition \textit{via} a catalytic vapor-liquid-solid method \cite{Wagner1964} (growth details were given in an earlier work \cite{Gentile2008}).  After growth, the Nws were dispersed on a heavily doped silicon substrate (to be used as a back gate) capped with a 300-nm-thick thermal oxide layer, and they were individually contacted with e-beam lithography by pairs of 120-nm-thick nickel electrodes deposited by e-beam evaporation. Prior to the deposition step the native oxyde was removed by a 5s dip into a solution of buffered hydrofluoridric acid (BHF). To enable the application of a heating electrical current, each nickel electrode was designed in the shape of a 500-nm-wide and 3-$\mu$m-long strip whose edges are connected to outer Cr(10 nm)/Au(65 nm) bonding pads \textit{via} progressively wider nickel metal lines defined in the same metal deposition step. 

\textbf{Temperature estimation}. An estimate of the temperature profile along the strip can be obtained from the solution of a one-dimensional heat equation under steady-state conditions \cite{Durkan1999}, \textit{i.e.} $$-\kappa\frac{d^2 \Delta T}{dx^2}+\frac{\kappa_{sub}}{t d}\Delta T=\rho J_{th}^2$$ where $\Delta T$ is the local incremental temperature (with respect to room temperature), $\rho$ is the electrical resistivity of nickel, $J_{th}$ is the heating  current density, $\kappa$ and $d$ ($\kappa_{sub}$ and $t$) are the thermal conductivity and thickness of the nickel strip (substrate surface oxide), respectively. With the parameters of our experiment ($\rho\approx 0.17\Omega\cdot\mu$m,  $\kappa=90$W/m$\cdot$ K, $\kappa_{sub} = 0.3$W/m$\cdot$ K, $d =300$nm, and $t=120$nm) we find that, away from the strip mid point,  $\Delta T$ falls off on a characteristic length ${\cal L}_{th}=\sqrt{t d\kappa/\kappa_{sub} } \sim 3.3 \mu$m, comparable to the total strip length. Thus an efficient Joule annealing can be expected whenever the misalignment between the position of the NW/strip contact and the strip mid point is smaller than a few hundred nm, \textit{i.e.} much less than ${\cal L}_{th}$. This condition was easily fulfilled in all the fabricated devices.

\begin{acknowledgments}
 
	We thank the technical staff of the PTA cleanroom, P. Payet-Burin, H. Blanc, G.Lapertot, C. Marin and A. Orsino for their help in device fabrication
and technical support. This work was supported by the Agence Nationale de la Recherche (ANR) through the ACCESS and COHESION projects and by
the European Commission through the Chemtronics program MEST-CT-2005-020513.

\end{acknowledgments}

\textbf{Supporting Information Available}: A video obtained from a sequence of consecutive SEM snapshots during the Joule-assisted silicidation of a Si NW device. Each snapshot was taken within about 20 s. This information is available free of charge \textit{via} the Internet at http://pubs.acs.org.

\bibliographystyle{achemso}

\bibliography{bibliography_nanowires}



\end{document}